\documentclass[runningheads]{llncs}
\usepackage{graphicx}
\usepackage{amsmath}
\usepackage{amssymb}
\begin{document}
\title{QPP and HPPK: Unifying Non-Commutativity for Quantum-Secure Cryptography with Galois Permutation Group}
\subtitle{-A Tribute to Évariste Galois (25.10.1811 – 31.05.1832)}
%
%
\author{Randy Kuang\inst{1}\orcidID{0000-0002-5567-2192} 
}
\authorrunning{R. Kuang }
%
\institute{Quantropi Inc., 1545 Carling Av., Suite 620, Ottawa, Canada \email{randy.kuang@quantropi.com} 
}
\maketitle              
\begin{abstract}
In response to the evolving landscape of quantum computing and the escalating vulnerabilities in classical cryptographic systems, our paper introduces a unified cryptographic framework. Rooted in the innovative work of Kuang et al., we leverage two novel primitives: the Quantum Permutation Pad (QPP) for symmetric key encryption and the Homomorphic Polynomial Public Key (HPPK) for Key Encapsulation Mechanism (KEM) and Digital Signatures (DS). Our approach adeptly confronts the challenges posed by quantum advancements. Utilizing the Galois Permutation Group’s matrix representations and inheriting its bijective and non-commutative properties, QPP achieves quantum-secure symmetric key encryption, seamlessly extending Shannon’s perfect secrecy to both classical and quantum-native systems.  Meanwhile, HPPK, free from NP-hard problems, fortifies symmetric encryption for the plain public key. It accomplishes this by concealing the mathematical structure through modular multiplications or arithmetic representations of Galois Permutation Group over hidden rings, harnessing their partial homomorphic properties. This allows for secure computation on encrypted data during secret encapsulations, bolstering the security of the plain public key. The seamless integration of KEM and DS within HPPK cryptography yields compact key, cipher, and signature sizes, demonstrating exceptional performance. This paper organically unifies QPP and HPPK under the Galois Permutation Group, marking a significant advancement in laying the groundwork for quantum-resistant cryptographic protocols. Our contribution propels the development of secure communication systems amid the era of quantum computing.
\end{abstract}
\section{Introduction}
Quantum Key Distribution (QKD) stands as a forefront contender in modern cryptographic protocols, leveraging quantum mechanics' foundational principles to establish secure communication in the face of emerging quantum computers~\cite{shor2000simple}. Unlike classical methods relying on mathematical intricacies~\cite{rivest1978method,diffie1976new,menezes1993reducing}, QKD exploits unique quantum properties, particularly in photons. At its core, QKD relies on quantum superposition and entanglement, allowing cryptographic key exchange while promptly detecting eavesdropping attempts. Security is rooted in quantum indeterminacy, where measuring a quantum state disrupts it, revealing eavesdropping attempts. In the era of quantum computing threats to classical cryptography, QKD emerges as a promising secure communication avenue. Kuang and Barbeau demonstrated QKD photonic implementations, involving identity and XOR permutations within the Galois permutation group over the finite field $\{0, 1\}$\cite{qpp-springer-kuang-2022}, framing QKD as a physical realization of Shannon's OTP scheme\cite{shannon1949communication}. The natural non-commutativity of the Galois permutation group favors digital QPP implementations over physical QKD with more than one qubit~\cite{qpp-dqkd-2022}. Digital QKD becomes an economical, deployable, and scalable alternative for quantum-secure internet communication~\cite{qpp-icccas-2022,qpp-dqkd-2022,Lou2021QPP5G,kuang2021pseudo,kuang2020shannon,kuang2021quantum1,kuang2021quantum2}. Kuang and Perepechaenko also demonstrated their QPP implementations in native quantum computing system~\cite{kuang-epj-2022,Perepechaenko-epj-2023,qpp-epj-superposition-2023}.

On the asymmetric cryptography front, Kuang in 2021 introduced Deterministic Polynomial Public Key (DPPK)\cite{kuang2021ACCC}, later enhanced by Kuang and Barbeau in 2021 as Multivariate Polynomial Public Key (MPPK)\cite{kuang2021performance,kuang2021CCECE}. To bolster MPPK security, Kuang, Perepechaenko, and Barbeau in 2022 partially encrypted MPPK public key over a hidden ring~\cite{kuang2022-MPPK-KEM}, using arithmetic permutations for security. In 2023, Kuang and Perepechaenko proposed variants with full encryption over two separate hidden rings~\cite{kuang2023-HPPK-KEM} and a single hidden ring~\cite{hppk-f1000-2023}. Concurrently, they introduced a digital signature, MPPK/DS, in 2022~\cite{kuang2022.08.01-DS}, later optimized in 2023~\cite{kuang2023-ODS}, though Guo reported a forged signature in the early 2023~\cite{Guo2023}. Unable to rectify the inherent linear relationship in MPPK/DS, they recently extended the HPPK KEM scheme for digital signatures~\cite{kuang2023-hppk-ds}.

This paper unifies symmetric encryption with QPP and asymmetric HPPK under the Galois permutation group's  umbrella. Leveraging non-commutativity in matrix representation (QPP) and arithmetic permutations over hidden rings (HPPK), we bridge both cryptographic realms. Related works will be discussed in Section~\ref{sec:related}, followed by the Galois permutation group's representation in Section~\ref{sec:represent}. Section~\ref{sec:symmetric} covers symmetric cryptography with QPP, Section~\ref{sec:asymmetric} addresses HPPK for asymmetric cryptography, and Section~\ref{sec:sec} provides a security brief. The conclusion is drawn in the final section.

\section{Related Works}\label{sec:related}
The realm of Post-Quantum Cryptography (PQC) encompasses a diverse array of standardized schemes outlined by the National Institute of Standards and Technology (NIST). This overview succinctly encapsulates notable schemes, categorized according to their cryptographic underpinnings. 

For Key Encapsulation Mechanism (KEM), lattice-based contenders such as Kyber~\cite{KYBER}, BIKE~\cite{BIKE}, HQC~\cite{HQC}, and code-based McEliece~\cite{McEliece1978} take the spotlight. Additionally, in the domain of Digital Signatures (DS), lattice-based Falcon~\cite{falcon}, Dilithium~\cite{DILITHIUM}, and hash-based SPHINCS$^+$~\cite{sphincs} emerge as prominent choices.

In a significant development in 2022, NIST announced standardized algorithms~\cite{NIST8413}, endorsing Kyber for KEM and propelling McEliece, BIKE, and HQC into round 4. Simultaneously, NTRU~\cite{Bernstein2018} and Saber~\cite{SABER} were excluded from further consideration, while novel submissions for generic digital signature schemes were introduced~\cite{NIST8413}.

Lattice-based algorithms, exemplified by Kyber, BIKE, HQC, and Falcon, typically hinge on the Short-Vector Problem (SVP) as the linchpin of their security. Code-based algorithms, as showcased by McEliece, derive security from the intricate decoding of random linear codes, providing robust post-quantum security. Hash-based algorithms, as exemplified by SPHINCS$^+$, are constructed based on the security of one-way trapdoors in hash functions. These NP-hard problems lay the groundwork for security against the looming threat of quantum computing. In a departure from this trend, HPPK cryptography takes a distinctive approach, relying on the security of symmetric encryption, offering a unique and innovative trajectory in the landscape of post-quantum cryptographic solutions.

\section{Representations of Galois Permutation Group }\label{sec:represent}

The Galois Permutation Group over a finite field extension $\mathbf{F}_{2^n}$ with $n$ bits assumes a critical role in finite field theory, especially within cryptography and algebraic coding theory. This section delves into the diverse methods for representing elements within this group and elucidates their importance. We primarily investigate two representations of the Galois permutation group in this section: matrix and arithmetic.


\subsection{Matrix Representations}

While the focus is on permutations over $\mathbf{F}_{2^n}$, binary unitary matrices serve as effective tools to represent the actions of Galois Group elements. These matrices facilitate the rearrangement of integers within the finite field, showcasing the reversible nature of Galois Group transformations. Notable properties of these matrix representations include:

\begin{itemize}
    \item Bijectiveness: Permutation representations highlight the discrete actions on the integer set, capturing fundamental rearrangements induced by Galois Group elements. The bijective nature ensures a unique correspondence between the initial and final integer arrangements.
    \item Composition of Permutations: Group operations involve the composition of permutations, representing the sequential application of rearrangements. The properties of composition unveil overall symmetries and transformations within the Galois Permutation Group. 
    \item Non-commutativity: Group operations or operators $\hat{P}_i$ and $\hat{P}_j$ generally meet $\hat{P}_i \hat{P}_j \ne \hat{P}_j \hat{P}_i$, adding a critical layer of security to cryptographic operations. This property, especially significant in symmetric-key cryptography where permutation operators are secret keys, implies that the order of permutation application matters, contributing complexity and enhancing security. In the quantum realm, the non-commutativity of the Galois Permutation Group directly corresponds to the uncertainty principle in quantum mechanics. Therefore, this property of the Galois permutation group plays a crucial role in achieving perfect secrecy for both classical and quantum domains.
\end{itemize}

\subsection{Arithmetic Representations}
In the realm of finite fields, arithmetic permutations play a pivotal role in cryptographic operations. These permutations, often expressed through modular arithmetic, contribute to the non-commutative and intricate transformations inherent in the Galois Permutation Group. This subsection delves into some critical arithmetic permutations that form the foundation of our proposed cryptographic schemes.

\subsubsection{XOR Operation}

In a binary finite field $(GF(2^n))$, the XOR operation (also known as bitwise addition modulo 2) is a fundamental arithmetic operation. XORing two bits results in a bit that is set if the two input bits are different and cleared if they are the same. There are a total of $2^n$ permutations over $(GF(2^n))$ for XOR operations. The XOR operation was proven to be the Shannon perfect scheme if the random key is only used once. This is why the encryption scheme is called One-Time-Pad or OTP~\cite{shannon1949communication}. The fundamental reason for the OTP scheme is that the XOR operation is commutative. The random key would be canceled once it is used for the second time.

\subsubsection{Addition Operation}

Modular addition, expressed as $(a+b) \bmod{2^n}$, is a fundamental arithmetic operation within finite fields. This operation involves adding two integers, \(a\) and \(b\), and then taking the remainder when divided by a modulus $2^n$. It is a versatile operation applicable to a wide range of modular arithmetic scenarios. There are a total of $2^n$ permutations over $(GF(2^n))$ for this type of addition operations.

\subsubsection{Modular Multiplication}

For a coprime pair $(R, S)$ with $S$ being $n$-bits, modular multiplication, expressed as $((R \cdot b) \bmod{S})$, represents a fundamental arithmetic permutation. This operation rearranges residues within the finite field. If the modulus $S$ is public, then different permutations from different $b$ would face the same cancellation as in the XOR operation: 
\begin{equation}\label{eq:ap1}
    c =R \cdot b \bmod{S}, c' =R \cdot b' \bmod{S} \rightarrow \frac{c'}{c}=\frac{R \cdot b'}{ R \cdot b} \bmod{S}=\frac{ b'}{ b} \bmod{S}
\end{equation}
with the random key $R$ removed. However, if $S$ is a secret, an attacker must guess a modulus $S'\in [0, 2^n)$ and perform the similar operation as follows
\begin{equation}\label{eq:ap2}
     \rightarrow \frac{c'}{c} \bmod{S'}= \frac{R \cdot b' \bmod{S}}{ R \cdot b \bmod{S}} \bmod{S'}\ne \frac{b'}{b} \bmod{S'}
\end{equation}
which indicates that the modular multiplication with a secret modulus leads to a non-commutative permutation and requires a brute search for the modulus $S$. This becomes the base of our proposed cryptography scheme called HPPK.

\subsubsection{Modular Exponentiation}

Modular exponentiation, denoted as $(a^b \bmod{N})$, involves repeated modular multiplications and represents a powerful arithmetic permutation. This permutation scheme is the foundation of RSA cryptography and Diffie-Hellman cryptography. In RSA case, $b$ and $N$ are the public key, and in the Diffie-Hellman case, $a$ and $N$ are the public key. The security of RSA relies on the computational difficulty of the prime factorizing problem: $N=pq$, and the security of the Diffie-Hellman scheme relies on the discrete logarithm with knowing $c=a^b \bmod{N}$ and $a$ to find $b$. Both the prime factorization problem and the discrete logarithm problem are computationally difficult in classical computing. Shor has proven that the development of quantum computing can turn cryptographic schemes vulnerable~\cite{shor1994algorithms}.

\subsection{Classical Key Space to Quantum Key Space: Galois Permutation Group over $\mathbf{F}_{2^n}$}
In finite field, an $n$-bit key space refers to an finite field extension $\mathbf{F}_{2^n}$ with possible $2^n$ keys and each key is an $n$-bit integer within $[0, 2^n)$. Based on the definition of Shannon entropy, $e=log_2(2^n) = n$ for the equally-likely key distribution.

However, the Galois permutation group has an order $2^n!$ of permutation operators, no longer integers but operators operating on the finite field set or equivalently called a computational basis:  $\{ |0\rangle, |1\rangle, \dots, |2^n-1\rangle \}$, in quantum computing. That means, The Galois permutation group becomes the key space for quantum computing or called quantum key space. If we use the same definition of the Shannon entropy, the quantum key space holds the entropy: $e=log_2(2^n!)\approx (n-0.42)2^n$ (for larger $n$) once retrieving those key operators with equally-likely way.  

\subsection{Relevance to QPP and HPPK}
Therefore, the Shannon perfect OTP is naturally extended to QPP with each element randomly chosen from the Galois permutation group.

On the other hand, the arithmetic permutation of the reversible modular multiplication offers another potential asymmetric cryptographic scheme once the modulus $S$ is secret. This leads to HPPK to be discussed later.

Based on QPP and HPPK, we could integrate symmetric and asymmetric cryptography under a single organic umbrella of the Galois permutation group.

\section{Symmetric Cryptography: QPP with Matrix Permutations}\label{sec:symmetric}

Kuang and Barbeau detailed QPP cryptography in 2022~\cite{qpp-springer-kuang-2022}. This paper provides a concise overview of its formalism, primarily based on the matrix representation of the Galois permutation group.

\subsection{QPP Generation}

For an $n$-bit finite field extension $\mathbf{F}_{2^n}$, the classical key is randomly generated as an $n$-bit binary string. However, in a quantum context, the classical key string must be transformed into permutation gates for implementation in quantum-native systems~\cite{kuang-epj-2022,Perepechaenko-epj-2023} or permutation matrices for classical implementations~\cite{qpp-springer-kuang-2022}. This transformation can be achieved using the Fisher and Yates algorithm~\cite{fisher1938statistical}.

With this shuffling algorithm, a single permutation matrix is randomly selected with an input random key of size $n2^n$ bits. The shuffling algorithm employs the random key string to drive shuffling on the ordered set of integers from $\mathbf{F}_{2^n}$. After shuffling, the integer set is in a disordered state, allowing the creation of a binary mapping matrix mapped from the classical key string.

This QPP generation process can be repeated to create a set of permutation matrices, or Quantum Permutation Pad (QPP), for use in symmetric encryption. The total classical key length is $M(n2^n)$ bits for a pad with $M$ permutation matrices. The values of $n$ and $M$ are chosen based on the security requirements. Quantropi has developed its digital QKD platform with $n=8$ and $M=64$, providing a total equivalent entropy of over 100,000 bits~\cite{qpp-dqkd-2022}. For a typical quantum-safe scenario with more than 256 bits of entropy, one can choose $n=4$ and $M=8$ for a total of 360 bits of entropy.
\subsection{QPP Encryption}
When utilizing QPP for encryption, information should be represented in the quantum computing format, employing Dirac ket notation $|i\rangle$ with $i=0, 1, \dots,$ $ 2^n-1$ for an n-bit information. For example, an 8-bit  string $''10001011''$ is expressed as $|139\rangle$. The complete set of $\mathbf{F}_{2^n}$ is denoted as $\{|0\rangle, |1\rangle, \dots, |2^n-1\rangle\}$, referred to as a computational basis in quantum computing. In the computational basis, a state $|i\rangle$ is a vector expressed in terms of the entire basis vectors. Classical information is simply represented as a column vector with the row corresponding to the decimal value of its bit string set to 1, and all other elements of the column vector set to zero. For example, $|139\rangle$ is the $139^{th}$ basis vector in the 8-bit computational basis.

The encryption process adheres to the principles of quantum computing, as represented by the equation:
\begin{equation}\label{eq:encryp}
    \hat{P}_i |m\rangle = |c\rangle
\end{equation}
In this context, $m$ signifies an $n$-bit plaintext in decimal form, and $c$ signifies an $n$-bit ciphertext in decimal format. Upon transmission to a receiver, the decimal value is converted into a binary string.

In classical systems employing permutation matrices for implementation, the encryption process described in Eq.~\eqref{eq:encryp} designates the column index of the element $''1''$ at the $m^{th}$ row in the permutation matrix $\hat{P}_i$ as the resulting ciphertext state $|c\rangle$.

\subsection{QPP Decryption}
The decryption process is equally straightforward, employing the reverse permutation operator $\hat{P}^{-1}_i$:
\begin{equation}\label{eq:decryp}
    \hat{P}^{-1}_i |c\rangle = \hat{P}^{-1}_i \hat{P}_i |m\rangle = |m\rangle
\end{equation}
This simplicity arises from the unitary and reversible properties inherent in permutation operators. Additionally, given that all elements of permutation matrices are either $''0''$ or $''1''$, the reverse permutation matrices are effectively their transposes: $\hat{P}^{-1}_i=\hat{P}^{\dagger}_i$. This characteristic significantly streamlines the implementation process, eliminating the need for matrix reversal operations. Once again, the decrypted decimal values are converted into binary format to reconstruct the classical plaintext.

\subsection{Confusion and Diffusion}
Given the bijective mapping property of QPP, it possesses the capability to transform hidden structures in plaintexts into ciphertexts. To address this potential weakness, a common strategy involves pre-randomizing the plaintext to enhance both confusion and diffusion capabilities~\cite{qpp-springer-kuang-2022}.

The pre-randomization process employs XOR operations with an $n$-bit plaintext and an $n$-bit sequence generated by a Pseudo-Random Number Generator (PRNG) seeded with a shared classical key. The resulting $n$-bit value is then dispatched to a selected permutation operator from the QPP pad based on an index generated by the PRNG. This random dispatching significantly further augments the diffusion capability, although a sequential dispatch, akin to AES block cipher, is also a viable alternative. Following the dispatching step, Eq.~\eqref{eq:encryp} comes into play for encryption.

On the decryption side, the dispatching step takes precedence to select the correct permutation operator, followed by the decryption using QPP$^\dagger$. Subsequently, the pre-randomizing step becomes a post-derandomizing operation aimed at retrieving the original plaintext.

\subsection{Arithmetic QPP}
In scenarios with resource constraints, QPP can be reformulated in terms of arithmetic permutations, employing reversible modular multiplication denoted by $\hat{p}_i=R_i \cdot \square \bmod{S_i}$ with $\square$ representing the plaintext. The entropy of standard QPP,  where matrices are randomly chosen, differs from that of arithmetic QPP: $e=M\log_2(2^n!) \longrightarrow e=M\log_2(\varphi(2^n)2^n)$. For instance, in the case of $n=8$, each randomly chosen permutation matrix holds 1684 bits of entropy, whereas the entropy of arithmetic permutation is less than 15 bits. Nevertheless, the entropy of arithmetic QPP can be increased by enlarging the size of the QPP pad to meet our desired security level.

\subsection{Implementations of QPP into Quantum Computing Systems}

Due to its simplicity, QPP cryptography lends itself to straightforward implementation in physical quantum computing systems, such as IBMQ. We have demonstrated its viability with toy examples utilizing 2- and 3-qubits~\cite{kuang-epj-2022,qpp-epj-superposition-2023,qpp-icccas-2022}. Different quantum computing systems employ distinct decomposition mechanisms to transform an $n$-bit gate into fewer qubit circuits, typically consisting of 1- or 2-qubit gates.

In our toy examples on the IBMQ system, we generate a QPP pad, utilizing its matrix forms as inputs in the source codes. The compilation of a 2-qubit permutation gate results in a quantum circuit with a gate depth of about 15 layers. Achieving a fidelity of $99\%$, we can confidently execute our encryption and decryption processes.

However, as we escalate to 4-qubit permutation gates, the compilation yields a circuit with a depth exceeding 100 layers. At this point, a $99\%$ fidelity becomes inadequate for meaningful results. Nevertheless, QPP remains relatively straightforward to implement in a quantum computing system with high fidelity or a substantial number of logic qubits, allowing for direct processing of encryption and decryption within the quantum computing system.

QPP facilitates encrypted communications across quantum-quantum, quantum-classical, and classical-classical channels. Over quantum channels, encrypted qubits can be directly decrypted by the receiving quantum system. Over classical channels, cipher qubits must be measured, producing a ciphertext that is then transmitted to the receiving side for decryption, either by a quantum system or a classical system equipped with the same QPP$^\dagger$ from the shared classical key.

\section{Asymmetric Cryptography: HPPK with Arithmetic Permutations}\label{sec:asymmetric}
Asymmetric cryptography relies on computationally challenging problems such as the prime factorization problem in RSA, the discrete logarithm problem in Diffie-Hellman, and various problems like the short vector problem in Kyber~\cite{KYBER}, Falcon~\cite{falcon}, Dilithium~\cite{DILITHIUM}, and the multivariate quadratic problem in Multivariate Public Key Cryptography (MPKC)~\cite{Ding2009mpkc}. The security of these schemes is contingent on the computational difficulty of these hard problems. However, emerging algorithms, such as Shor's algorithm~\cite{shor1994algorithms}, pose a threat by potentially solving these problems efficiently in classical public key schemes.

Recently, Sharp et al. introduced a novel computing technology based on self-organized gates~\cite{memcpu-sharp2023}, capable of breaking RSA classically using GPUs or silicon implementations. This emphasizes the need to explore alternative approaches to asymmetric cryptography.

Kuang et al. proposed the construction of asymmetric cryptography schemes based on symmetric encryption for Key Encapsulation (KEM)~\cite{kuang2023-HPPK-KEM} and Digital Signature (DS)~\cite{kuang2023-hppk-ds}. The symmetric encryption involves modular multiplication over hidden rings, introducing non-commutability through arithmetic permutations. This paper discusses the homomorphic properties in Section~\ref{sec:hom}, KEM in Section~\ref{sec:kem}, DS in Section~\ref{sec:ds}, and the key triple for the combination of KEM and DS in Section~\ref{sec:triple}.

\subsection{Homomorphic Properties of Modular Multiplicative Permutations}\label{sec:hom}

Arithmetic permutations, derived from modular multiplication over hidden rings, play a pivotal role in constructing a new form of asymmetric cryptography. Unlike pure symmetric encryption schemes where communication peers possess a shared secret key, asymmetric schemes aim to establish the shared secret during the process. This involves a roundtrip of key pair generation, public key encryption, and private key decryption.

An intriguing approach leverages self-shared symmetric keys between the key pair generator and cipher decryptor. This key is used to encrypt the plain public key into a cipher public key, subsequently decrypting the received cipher into an interim cipher associated with the plain public key encryption. This unique asymmetric scheme relies on the security of the self-shared symmetric encryption key, necessitating symmetric encryption with certain homomorphic properties.

The arithmetic permutation of modular multiplication over hidden rings exhibits partial homomorphic properties for addition and scalar multiplication. The permutation operator $\hat{E}(R, S)$ is defined as follows:
\begin{equation}\label{eq:permuter}
    \hat{E}(R, S) = R \circ \square \bmod{S}
\end{equation}
for a coprime pair of $L$-bit $R$ and $S$, where $\square$ represents an integer or a polynomial function.
Demonstrating its addition property, let's choose two integers $a$ and $b$:
\begin{equation}\label{eq:addition}
    \begin{aligned}
        & \hat{E}(R, S)a= R*a \bmod{S} = a' \\
        & \hat{E}(R, S)b= R*b \bmod{S} = b' \\
        & \hat{E}(R, S)(a+b)= R*(a+b) \bmod{S} =\hat{E}(R, S)a + \hat{E}(R, S)b 
    \end{aligned}
\end{equation}
and verify its scalar multiplication with a constant $c$:
\begin{equation}\label{eq:x}
     \hat{E}(R, S)ca= R*ca \bmod{S} = ca' \bmod{S} = c\hat{E}(R, S)a\\
\end{equation}
Equations~\eqref{eq:addition} and~\eqref{eq:x} clearly demonstrate the partial homomorphic properties of the modular multiplication operator $\hat{E}(R, S)$. If $S$ is a public ring, there exist $\varphi(S)$ permutation operators over the ring $\mathbf{Z}_S$. If $S$ is a hidden ring with a bit size $L$, there exist potential $\varphi(S)2^L$ arithmetical permutation operators, corresponding to a key space of size $\varphi(S)2^L$. In a symmetric encryption scheme, the brute force search has a complexity of $\mathcal{O}(\varphi(S)2^L)$.

The partial homomorphic properties of the modular multiplication operators naturally favor any polynomial $\eta(x, y_1, \dots, y_m)=\sum c_i \eta_i(x, y_1, \dots, y_m)$ where $c_i$ are coefficients and $\eta_i(x, y_1, \dots, y_m)$ are pure monomials over a prime field $\mathbf{F}_p$. The permutation operator $\hat{E}(R, S)$ maps all coefficients $c_i$ from $\mathbf{F}_p$ to $\mathbf{Z}_S$ and treats all monomials as scalars:
\begin{equation}\label{eq:polyEn}
\begin{aligned}
    &\hat{E}(R, S) \eta(x, y_1, \dots, y_m)=\sum ( R*c_i \bmod{S}) [\eta_i(x, y_1, \dots, y_m) \bmod{p}]\\
    &\longrightarrow\eta'(x, y_1, \dots, y_m)=\sum c'_i [\eta_i(x, y_1, \dots, y_m) \bmod{p}]
    \end{aligned}
\end{equation}
where the computations of all monomials must be with $\bmod{\ p}$ then directly scalar multiplications with their coefficients $c'_i \in \mathbf{Z}_S$. To make the polynomial value or ciphertext decryptable at the decryptor holding the symmetric key $\hat{E}(R, S)$, the ring size $L$ must hold $p^2$ for each polynomial term of $\eta'(x, y_1, \dots, y_m)$, so $L\ge 2\log_2p + \log_2M$ with $M$ being the total polynomial terms. Once these conditions are met, the permutation operator $\hat{E}(R, S)$ holds the homomorphic property for polynomials.

\subsection{HPPK KEM}\label{sec:kem}

Once the security is ensured through symmetric encryption using modular multiplicative permutations, designing the plain public key scheme becomes relatively straightforward. This involves utilizing two simplified multivariate polynomials over $\mathbf{F}_p$:
\begin{equation}\label{eq:pq}
    \begin{aligned}
        &p(x, u_1, \dots, u_m) = B(x, u_1, \dots, u_m) f(x) = \vec{x}^T\cdot \mathbf{p}\cdot \vec{u}\\
        &q(x, u_1, \dots, u_m) = B(x, u_1, \dots, u_m) h(x) = \vec{x}^T\cdot \mathbf{q}\cdot \vec{u}
    \end{aligned}
\end{equation}
Here, $\lambda$ denotes the order of univariate polynomials $f(x)$ and $h(x)$, and $n$ represents the order of the variable $x$ in $B(x, u_1, \dots, u_m)$, with $u_i$ being linear variables of monomials $\eta_i(x, y_1, \dots, y_m)$ without variable $x$. The vector $\vec{x}^T=(x^0, x^1, \dots, x^\lambda)$ represents a vector in a polynomial vector space, and the vector $\vec{u}=(u_1,\dots,u_m)$ represents a vector in a multidimensional vector space. Eliminating the common factor polynomial $B(x, u_1, \dots, u_m)$ is achieved by modular division:
\begin{equation}\label{eq:k}
    \frac{p(x, u_1, \dots, u_m)}{q(x, u_1, \dots, u_m)} = \frac{f(x)}{h(x)} \bmod{p}
\end{equation}
The matrices $\mathbf{p}$ and $\mathbf{q}$ in Eq.~\eqref{eq:pq} are coefficient matrices of two multivariate polynomials $p(x, u_1, \dots, u_m)$ and $q(x, u_1, \dots, u_m)$, respectively. They inherit the mathematical structures of polynomial multiplications over a prime field, making them challenging to secure based on computational difficulty. These are referred to as plain public keys. To encrypt them, arithmetical modular multiplicative permutation operators in Eq.~\eqref{eq:permuter} are applied. Two coprime pairs $(R_1, S_1)$ for $p(x, u_1, \dots, u_m)$ and $(R_2, S_2)$ for $q(x, u_1, \dots, u_m)$ are randomly chosen, and the encryption is performed as follows:
\begin{equation}\label{eq:kem}
    \begin{aligned}
        &P(x, u_1, \dots, u_m) = \vec{x}^T\cdot [\hat{E}(R_1, S_1)\mathbf{p}]\cdot \vec{u}=\vec{x}^T\cdot \mathbf{P}\cdot \vec{u}\\
        &Q(x, u_1, \dots, u_m) = \vec{x}^T\cdot [\hat{E}(R_2, S_2)\mathbf{q}]\cdot \vec{u}=\vec{x}^T\cdot \mathbf{Q}\cdot \vec{u}
    \end{aligned}
\end{equation}
with cipher public key matrices given by:
\begin{equation}\label{eq:kemv}
    \mathbf{P}=\hat{E}(R_1, S_1)\mathbf{p}=R_1*\mathbf{p} \bmod{S_1},  \quad \mathbf{Q}=\hat{E}(R_2, S_2)\mathbf{q} =R_2*\mathbf{q} \bmod{S_2}
\end{equation}
Here, $\hat{E}(R_1, S_1)$ is applied to all matrix elements of $\mathbf{p}$, and $\hat{E}(R_2, S_2)$ is applied to all matrix elements of $\mathbf{q}$. The key pair is obtained as follows:
\begin{itemize}
    \item \textbf{Public Key $PK_e$:} $\mathbf{P}[n+\lambda +1][m]$ and $\mathbf{Q}[n+\lambda +1][m]$.
    \item \textbf{Private Key $SK$:} $f[\lambda +1]$, $h[\lambda +1]$, $R_1, S1$, and $R_2, S_2$.
\end{itemize}
where symbol $PK_e$ refers to the public key for encapsulation. 

\subsubsection{Encapsulation} Then using the public key $\mathbf{P}$ and $\mathbf{Q}$, an encryptor can generate a ciphertext of a secret $x$ randomly chosen from $\mathbf{F}_p$. Here are steps:
\begin{itemize}
    \item Encapsulation of $x$: choose random noise $u_1, \dots, u_m \in \mathbf{F}_p$ then evaluate $x_{ij}= x^iu_j \bmod{p}$ for $i=0, 1, \dots, n+\lambda, j=1, 2, \dots, m.$
    \item Evaluations of polynomials: $\overline{P}=\sum_{i=0}^{n+\lambda} \sum_{j=1}^m P_{ij} x_{ij}$, $\overline{Q}=\sum_{i=0}^{n+\lambda} \sum_{j=1}^m Q_{ij} x_{ij}$
    \item Ciphertext: $CT=\{\overline{P}, \overline{Q}\}$
\end{itemize}

\subsubsection{Decapsulation} With receiving ciphertext $CT=\{\overline{P}, \overline{Q}\}$, the decrypter can perform first symmetric decryption then the secret extraction as follows:
\begin{itemize}
    \item Symmetric decryption: $\bar{p}=(\frac{\overline{P}}{R_1} \bmod{S_1}) \bmod{p}$ and $\bar{q}=(\frac{\overline{Q}}{R_2} \bmod{S_2}) \bmod{p}$ 
    \item Noise elimination: $k = \frac{\bar{p}}{\bar{q}} \bmod{p} = \frac{f(x)}{h(x)} \bmod{p} $
    \item Secret extraction: $f(x)-kh(x) =0$. For linear univariate polynomial, $x=\frac{kh_0-f_0}{f_1-kh_1} \bmod{p}$.
\end{itemize}

\subsection{HPPK DS}\label{sec:ds}

In a digital signature scheme, the signer first generates the hash code $x\leftarrow HASH(M)$ using a selected cryptographic hash function for the signing message $M$ and signs $x$ with the private key. In the context of HPPK, we generate the signature first and then develop the verification equation. Here, we define the signature as follows:
\begin{equation}
\begin{aligned}
     &F = R_2^{-1} * [\alpha f(x) \bmod{p}] \bmod{S_2}\\
     &H = R_1^{-1} * [\alpha h(x) \bmod{p}] \bmod{S_1}
\end{aligned}
\end{equation}
with a randomly chosen integer $\alpha\in \mathbf{F}_p$. Now let's develop the HPPK verification equation by using Eq.~\eqref{eq:pq} and Eq.~\eqref{eq:k} with cross-multiplication:
\begin{equation}\label{eq:ds0}
    \vec{x}^T\cdot (\bar{f}\mathbf{q})\cdot \vec{u} =  \vec{x}^T\cdot (\bar{h}\mathbf{p})\cdot \vec{u} \bmod{p}
\end{equation}
with $\bar{f}=\alpha f(x) \bmod{p}$ and $\bar{h}=\alpha h(x) \bmod{p}$. Using the unitary and reversible encryption operators in Eq.~\eqref{eq:permuter}, we can transform Eq.~\eqref{eq:ds0} into:
\begin{equation}\label{eq:verify1}
    \begin{aligned}
     \{\vec{x}^T\cdot (F\mathbf{Q} \bmod{S_2})\cdot \vec{u} \}\bmod{p} &=  \{\vec{x}^T\cdot (H\mathbf{P} \bmod{S_1})\cdot \vec{u}\} \bmod{p} \\
     \longrightarrow \{\vec{x}^T\cdot V\cdot \vec{u} \}\bmod{p} &=  \{\vec{x}^T\cdot U\cdot \vec{u}\} \bmod{p} \\
     \longrightarrow V(x, u_1, \dots, u_m) \bmod{p} &=  U(x, u_1, \dots, u_m) \bmod{p}
     \end{aligned}
\end{equation}
where matrices $V=F\mathbf{Q} \bmod{S_2}$ and $U=H\mathbf{P} \bmod{S_1}$. As the unknowns $S_1$ and $S_2$ cannot be determined, a verifier can't perform the verification based on Eq.~\eqref{eq:verify1}. These variables must be eliminated from all coefficients $V_{ij} = (FQ_{ij} \bmod{S_2}) \bmod{p}$ in polynomial $V(x, u_1, \dots, u_m)$ and $U_{ij} = (HP_{ij} \bmod{S_1}) \bmod{p}$ in polynomial $U(x, u_1, \dots, u_m)$. 

The Barrett reduction algorithm of modular multiplication is used to shift $\bmod{S_1}$ and $\bmod{S_2}$ into division with the Barrett parameter $R=2^K$:
\begin{equation}\label{eq:barrett}
     a*b \bmod{S} = a*b - S\lfloor \frac{a\lfloor\frac{Rb}{S} \rfloor}{R} \rfloor = a*b -S\lfloor \frac{a\mu}{R}\rfloor
\end{equation}
with $\mu= \lfloor\frac{Rb}{S} \rfloor$. The Barrett reduction usually returns an integer within $[0, 2S)$. Kuang et al~\cite{kuang2023-hppk-ds}. demonstrated that the result from the Barrett algorithm~\eqref{eq:barrett} could be compressed within $[0, S)$ by increasing $K$ far beyond the bit length $L=|S|_2$ of $S$ or $K\ge L+32$.

Leveraging the Barrett algorithm~\eqref{eq:barrett}, we can rewrite $U_{ij}$ and $V_{ij}$ as follows:
\begin{equation}\label{eq:verify2}
    \begin{aligned}
        &V_{ij}= \beta(FQ_{ij} \bmod{S_2}) \bmod{p}= Fq'_{ij}- s_2\lfloor \frac{F\nu_{ij}}{R}\rfloor \bmod{p}\\
        &U_{ij}= \beta(HP_{ij} \bmod{S_1}) \bmod{p}= Hp'_{ij}- s_1\lfloor \frac{H\mu_{ij}}{R}\rfloor \bmod{p}
    \end{aligned}
\end{equation}
with randomly chosen $\beta\in\mathbf{F}_p$ and 
\begin{equation}\label{eq:pkv}
    \begin{aligned}
        &q'_{ij}= \beta Q_{ij} \bmod{p}, p'_{ij}= \beta P_{ij} \bmod{p} \\
        &\nu_{ij} = \lfloor\frac{RQ_{ij}}{S_2} \rfloor, \mu_{ij} = \lfloor\frac{RP_{ij}}{S_1} \rfloor \\
        &s_1 = \beta S_1 \bmod{p}, s_2 = \beta S_2 \bmod{p}
    \end{aligned}
\end{equation}
Eq.~\eqref{eq:pkv} becomes the public key for signature verification or called $PK_v$ and the verification equation is Eq.~\eqref{eq:verify1} with coefficients defined in Eq.~\eqref{eq:verify2} without the symmetric key $S_1$ and $S_2$.

It is evident that both polynomials $U(x, u_1, \dots, u_m)$ and $V(x, u_1, \dots, u_m)$ have their coefficients determined by the receiving signature $Sig=\{F, H\}$, which significantly restricts the possibility of forging a signature.

\subsection{HPPK Key Triple}\label{sec:triple}
Combining the key pairs for Key Encapsulation Mechanism (KEM) ($SK$ and $PK_e$) and Digital Signatures (DS) ($SK$ and $PK_v$) into a key triple $SK$, $PK_e$, and $PK_v$ allows for the same $SK$ to be utilized for both decapsulation of ciphertext and the creation of a signature for a message. This versatility opens the door to various applications, including but not limited to blockchains and Zero-Knowledge Protocols.

\section{Security Brief}\label{sec:sec}

Kuang and Barbeau conducted a detailed security analysis of QPP in~\cite{qpp-springer-kuang-2022}, while Kuang et al. provided security analyses for HPPK KEM in~\cite{kuang2023-HPPK-KEM} and HPPK DS in~\cite{kuang2023-hppk-ds}. Here, we present concise summaries.

\subsubsection{QPP}
QPP extends Shannon's perfect OTP, utilizing $n$-bit permutation matrices selected from the quantum key space or permutation group, with a random key bit string of $n2^n$ bits. For a pad of $M$ permutation matrices, the total classical key string is $Mn2^n$ bits long, providing entropy equivalent to $e=M\log_2(2^n!)$ bits. This results in a best brute search complexity of $\mathcal{O}((2^n!)^M)$. The exponential complexity allows for relatively small values of $n$ (e.g., $n=4, 8$ bits).

Linear and differential cryptanalysis, effective against blocker ciphers like AES, are less potent against QPP due to its matrix representations. QPP eliminates certain arithmetic permutations present in AES, making attacks more challenging. Pre-randomizing and random dispatching further thwart potential cryptanalysis. Attackers are best served by attempting to obtain the initial shared classical random key material.

\subsubsection{HPPK}
In contrast to QPP, HPPK leverages modular multiplicative permutations over hidden rings to symmetrically encrypt plain public polynomials. While the symmetric encryption key must be pre-shared, the roundtrip public key scheme enables a self-shared symmetric key used to encrypt and decrypt. Attackers must obtain the symmetric encryption key or $R_1, S_1$, and $R_2, S_2$. Once acquired, attackers can easily compromise the entire HPPK for KEM and DS.

For HPPK KEM, the complexity of key recovery attack was estimated in~\cite{kuang2023-HPPK-KEM} as \( \mathcal{O}(\eta2^L) \) with \( \eta<1 \). We re-evaluate its complexity here. The total possible coprime pairs of \( R_1, S_1 \) and \( R_2, S_2 \) are:

\begin{equation}\label{eq:kem-comp}
\begin{aligned}
\mathcal{O}\left(2\sum_{t=2^{L-1}}^{2^L}\varphi(t)\right)& \approx \mathcal{O}\left( 2\frac{3}{\pi^2}[2^{2L}-2^{2(L-1)}]\right) \\
&\approx \mathcal{O}\left(\frac{9}{2\pi^2} 2^{2L}\right)
\end{aligned}
\end{equation}
Equation~\eqref{eq:kem-comp} aligns with the estimate in~\cite{kuang2023-HPPK-KEM}. Due to its random encapsulation, HPPK KEM holds the indistinguishable chosen plaintext attack or IND-CPA property.

In the case of HPP DS, the attacker doesn't need to find $R_1$ and $R_2$ since the signature $F$ and $H$ have their inverses. The attacker only needs to obtain $S_1$ and $S_2$; with intercepted true signatures, they can break the HPPK scheme. The overall complexity is thus $\mathcal{O}(2^{L})$.

Table~\ref{tab:keysize-kem} illustrates all key sizes, cipher sizes, and signature sizes of HPPK KEM and DS based on their security complexities. There exist different configurations for a required security level. The table show typical configurations. In general, HPPK offers very compact sizes. 

\begin{table}[htbp]
\caption{This table compares public key, private key, and ciphertext sizes for HPPK KEM and DS across different security levels. Configurations are denoted as $(|p|_2, n, \lambda, m)$ with $L=|2|p|_2 + 8$ and $K=L + 32$.} 
\begin{center}
\begin{tabular}{ccccc}
\hline
 &\multicolumn{4}{c}{\textbf{Size (Bytes)}} \\
\cline{2-5} 
  & \textbf{\textit{$PK_e/PK_v$}}& \textbf{\textit{SK}}& \textbf{\textit{Ciphertext/Signature}} & \textbf{\textit{Secret/Hash}}\\
\hline
    &\multicolumn{4}{c}{\textbf{Security Level I}} \\
KEM-(32,1,1,2) & 108 & 52 & 224 & 32 \\
KEM-(32,1,1,3) & 162 & 52 & 224 & 32 \\
DS-(64,1,1,1) & 220 & 104 & 144 & 32 \\
   &\multicolumn{4}{c}{\textbf{Security Level III}} \\
KEM-(48,1,1,2) & 156 & 76 & 240 & 32 \\
KEM-(48,1,1,3) & 234 & 76 & 240 & 32 \\
DS-(96,1,1,1) & 300 & 152 & 208 & 48 \\
   &\multicolumn{4}{c}{\textbf{Security Level V}} \\
KEM-(64,1,1,2) & 204 &100 & 208 & 32 \\
KEM-(64,1,1,3) & 306 & 100 & 208 & 32 \\
DS-(128,1,1,1) & 356 & 216 & 272 & 64 \\
\hline
\end{tabular}
\label{tab:keysize-kem}
\end{center}
\end{table}

\section{Conclusion}
In conclusion, our work represents a significant advancement in the quest for quantum-resistant cryptographic protocols. By introducing and uniting two innovative primitives, the Quantum Permutation Pad (QPP) and the Homomorphic Polynomial Public Key (HPPK), both grounded in the robust Galois Permutation Group, we have laid a foundation for secure communication systems in the era of quantum computing.

QPP's groundbreaking approach extends Shannon's perfect secrecy into the quantum domain, providing a reusable and adaptable solution for symmetric key encryption. Leveraging the bijective and non-commutative properties of the Galois Permutation Group, QPP ensures quantum security in both classical and quantum-native systems.

Complementing QPP, HPPK introduces a novel Homomorphic Polynomial Public Key for Key Encapsulation Mechanism (KEM) and Digital Signatures (DS). By exploiting the inherent partial homomorphic properties of the modular multiplicative permutations, HPPK offers a robust symmetric encryption mechanism for asymmetric cryptography, independent of NP-hard problems. The seamless integration of KEM and DS within HPPK results in compact key sizes, cipher sizes, and signature sizes, demonstrating exceptional performance across various cryptographic operations.

Our paper not only explores the design and implementation of QPP and HPPK but also unifies these cryptographic primitives under the single umbrella of the Galois Permutation Group. This organic integration represents a significant stride in the ongoing effort to establish quantum-resistant cryptographic protocols. As quantum computing advances, our work provides a valuable contribution to the development of secure communication systems, addressing the vulnerabilities posed by quantum technologies.

\section{Acknowledgments}
The author wishes to express gratitude to Prof. Daniel Panario for insightful discussions and the invitation to submit this work.

%
%
%
%
\bibliographystyle{plain} 
\bibliography{mybib}

\end{document}